\documentclass[aps,pra,groupedaddress,showpacs,twocolumn,longbibliography,10pt]{revtex4-1}
\usepackage[utf8]{inputenc}
\usepackage{graphicx}
\usepackage[export]{adjustbox}
\usepackage{textcomp}
\usepackage{hyperref}
\usepackage{amsmath}
\usepackage{bbold}
\usepackage{natbib}
\usepackage{units}
\usepackage[normalem]{ulem}
\usepackage{color}
\usepackage{ifthen}
\usepackage{afterpage}
\usepackage{xspace}
\usepackage{xr}
\usepackage{textcomp}
\usepackage{hyperref}
\usepackage[dvipsnames]{xcolor}

\newcommand{\SB}{\text{memory basis}\xspace}

\newcommand{\IB}{\text{interface basis}\xspace}

\renewcommand{\sup}[1]{(see Supplementary\xspace #1)}
\definecolor{green}{rgb}{0.0, 0.5, 0.0}

\newcommand{\BFigure}[2] {(Fig. \ref{#1}#2)}

\newcommand{\bra}[1]{\langle #1 |} %
\newcommand{\ket}[1]{\ensuremath{{|#1\rangle}}} %

\newcommand{\mymath}[1]{\ensuremath{#1}\xspace} 

\newcommand{\iu}{{i\mkern1mu}}

\newcommand{\Rb}{$^{\text{87}}\text{Rb }$}
\def \DOne {\mymath{\text{D}_{1}}}

\newcommand{\Fmf}[2]{\mymath{\left| F{=}#1,\, m_F{=}#2 \right\rangle}}
\newcommand{\FPmf}[2]{\mymath{\left| F'{=}#1,\, m_F{=}#2 \right\rangle}}
\newcommand{\QubitState}{\Fmf{2}{{\pm} 1}}

\newcommand{\CstState}{\left\{\Fmf{1}{{-}1},\Fmf{2}{1}\right\}}
\newcommand{\Fidelity}{\ensuremath{\mathcal{F}}\xspace}

\newcommand{\citel}[1][]{%
	\red{\ifthenelse{\equal{#1}{}}{[?]}{[#1]}}%
}

\newcommand{\red}[1]{{\color{red}#1}}

\newcommand{\CiteLongCoherenceSystems}{\cite{Bar-Gill2013,Langer2005,Maurer1283,Steger1280,zhong2015optically,yang2016coherence}\xspace}
\newcommand{\CiteFutureQNetworks}{\cite{razavi2009quantum,kimble2008quantum}\xspace}
\newcommand{\CiteLightMatterInterfaces}{ \cite{Gouraud2015,Sayrin2015, sprague2014,julsgaard2004,choi2008,clausen2012}\xspace}
\newcommand{\CiteLongMemories}{\cite{PhysRevLett.111.240503,riedl2012bose}\xspace}

\newcommand \figureOne {
\begin{figure}[!t]
	\includegraphics[width=0.95\columnwidth]{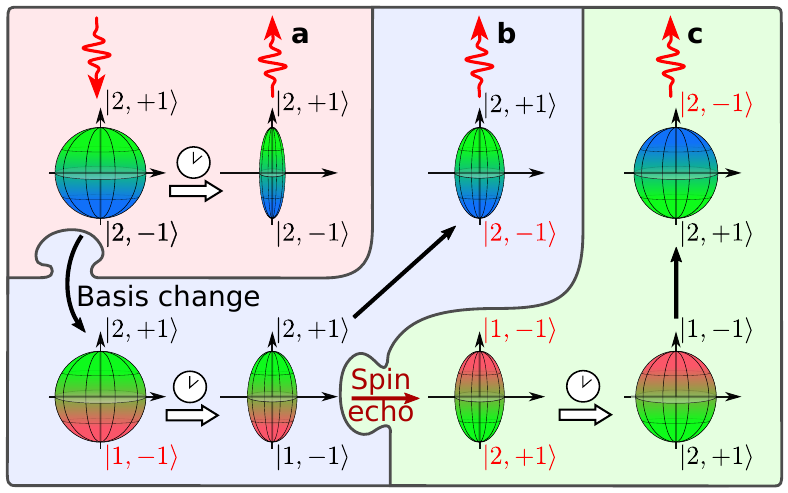}
\caption{
A photonic qubit (red wavy arrow pointing downwards) is mapped onto a superposition of two Zeeman substates $\Fmf{2}{{\pm}1}$ of the \Rb ground-state manifold by means of a stimulated Raman adiabatic passage.
Three experimental protocols are employed:
all experiments start by mapping the photonic qubit onto the atom and end by recreating the photon (upward pointing red wavy arrow).
\textbf{a,} In an elementary store-and-retrieve experiment (red area) the qubit dephases within hundreds of microseconds.
The dephasing is dominantly caused by magnetic field fluctuations.
This is illustrated as the deformation of the Bloch sphere which represents the qubit.
The labels \ket{F, m_F} denote the qubit basis states.
\textbf{b,} By temporarily mapping the qubit to a \textit{memory basis} which is less sensitive to those fluctuations, the rate of dephasing can be drastically reduced (blue area).
This allows for a total storage time on the order of tens of milliseconds.
A basis change is indicated by the changing color code of the Bloch spheres.
\textbf{c,} By additionally applying a spin-echo pulse (green area), residual dephasing mechanisms can be partly reverted leading to a total storage time beyond \unit[100]{ms}.
}
\label{puzzlepic}
\end{figure}
}

\newcommand \figureTwo {
\begin{figure}[!t]
\includegraphics[width=0.95\columnwidth]{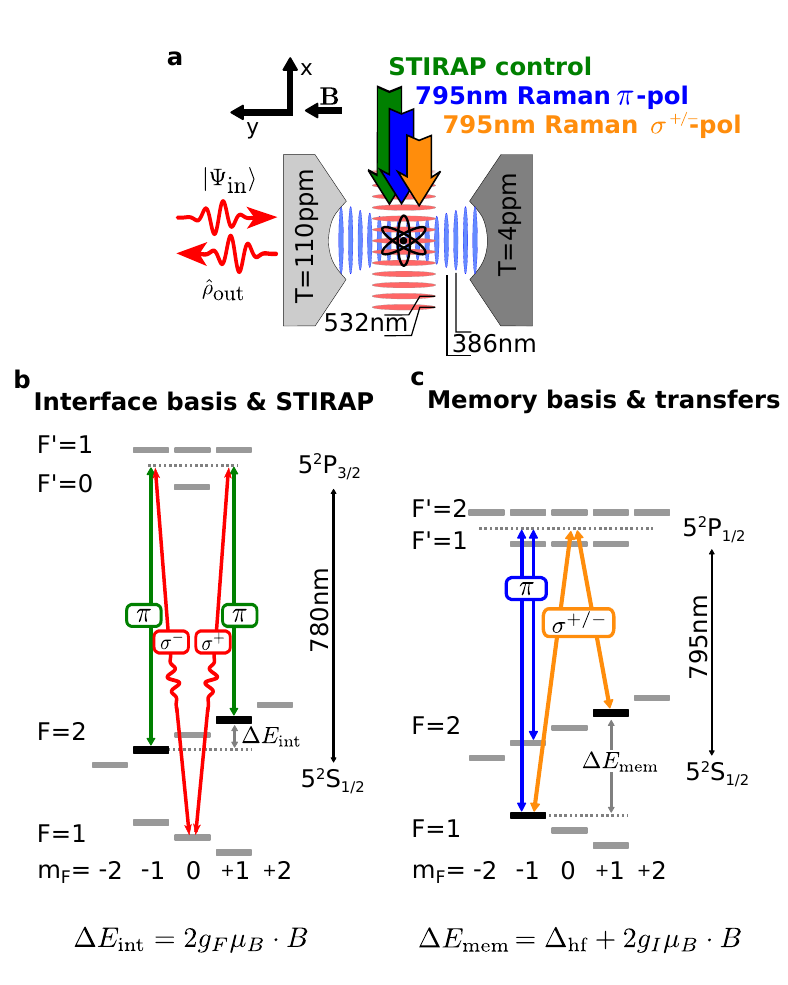}
\caption{
Experimental setup and level schemes for the single-atom quantum memory.
\textbf{a,} The atom is confined in a high-finesse optical cavity by a red-detuned standing-wave dipole trap and a blue-detuned intracavity dipole trap. The cavity has asymmetric mirror transmissions T resulting in intracavity photons predominately leaving the cavity through the outcoupling mirror. Incoming photons also enter through this mirror.
\textbf{b,} Level scheme for photon storage: the atom is initially pumped
to $\Fmf{1}{0}$. 
The incoming photon, whose polarisation can be decomposed in the $\sigma^-$ and $\sigma^+$ basis (red arrows) is resonant with the cavity. 
A control laser impinging perpendicular to the cavity (green arrows) transfers the photonic qubit to a superposition of atomic states $\QubitState$ via a stimulated Raman adiabatic passage.
\textbf{c,} By means of a Zeeman-state-selective Raman transfer using two $\pi$-polarised beams (blue arrows) the \IB can be mapped to the \SB $\CstState$, which is less susceptible to magnetic field fluctuations, and vice versa.
By using $\sigma^+/ \sigma^-$ polarised light (orange arrows), the population can be swapped, allowing to perform a spin echo.
}
\label{schematics}
\end{figure}
}

\newcommand \figureThree{
\begin{figure}[!t]
\includegraphics[width=1\columnwidth]{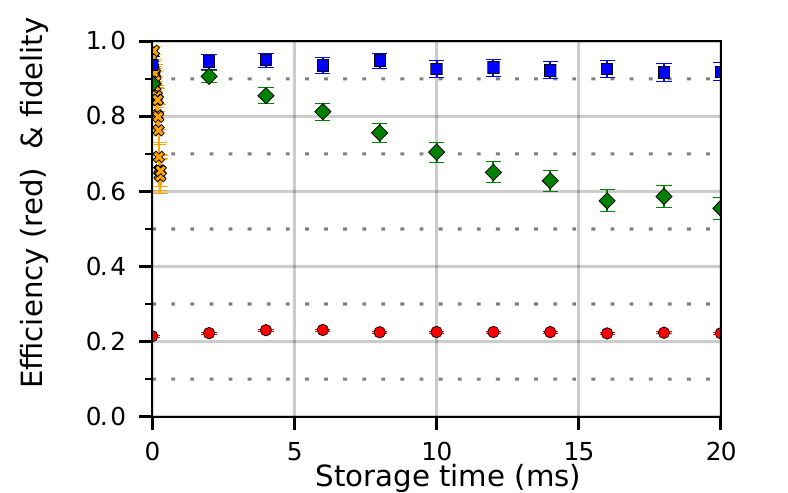}
 \caption{
Results for a memory experiment with mapping and remapping to and from the \SB, respectively.
Total memory efficiency (red circles) and average fidelity of the retrieved quantum state for linear (green diamonds) and circular (blue squares) input polarisations.
The average linear fidelity is determined by storing and retrieving horizontal, vertical, diagonal and anti-diagonal photons individually and creating the average of those fidelities.
For the average circular fidelity, the fidelities of left- and right-circular photons are measured individually and combined to an average.
For comparison, the decay of the fidelity for linear polarised input photons for a memory without mapping to the \SB is shown (yellow crosses).
The error bars represent the 95\% confidence intervals of the statistical uncertainties.
}
\label{cstplot}
\end{figure}
}

\newcommand \figureFour{
\begin{figure}[t]
\includegraphics[width=1\columnwidth]{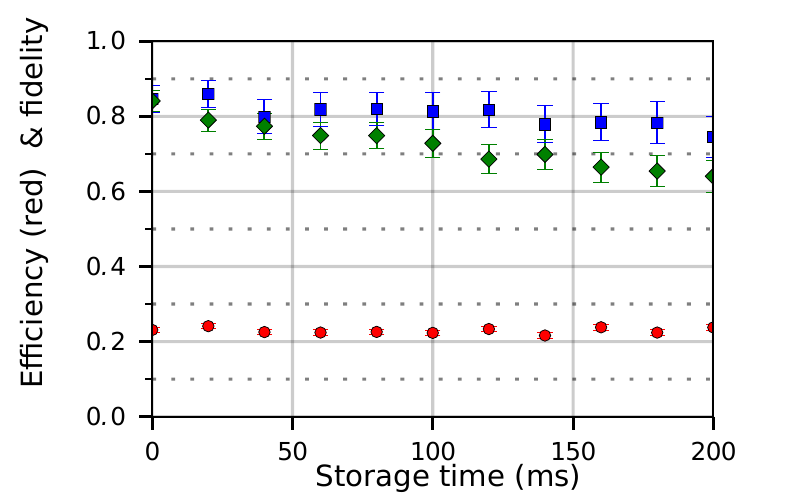}
 \caption{
Results for a memory experiment with the application of a spin-echo technique.
Total memory efficiency (red circles) and average fidelity of the retrieved quantum state for linear (green diamonds) and circular (blue squares) input polarisations.
The average fidelities are determined like in Figure 3.
The error bars represent the 95\% confidence intervals of the statistical uncertainties.
}
\label{seplot}
\end{figure}
}

\begin{document}
\title{Decoherence-protected memory for a single-photon qubit}
\author{M. Körber}
\email{matthias.koerber@mpq.mpg.de}
\author{$^\dagger$ O. Morin}
\thanks{M.K. and O.M. contributed equally}
\author{S. Langenfeld}
\author{A. Neuzner}
\thanks{Present Address: OHB System AG Weßling Germany}
\author{S. Ritter}
\thanks{Present Address: TOPTICA Photonics AG Graefelfing Germany}
\author{G. Rempe}
\affiliation{Max-Planck-Institut f\"{u}r Quantenoptik, Hans-Kopfermann-Strasse 1, 85748 Garching, Germany}

\maketitle
{
\small\noindent\textbf{
%\
The long-lived, efficient storage and retrieval of a qubit encoded on a photon is an important ingredient for future quantum networks \CiteFutureQNetworks.
Although systems with intrinsically long coherence times have been demonstrated \CiteLongCoherenceSystems, the combination with an efficient light-matter interface \CiteLightMatterInterfaces remains an outstanding challenge.
In fact, the coherence times of memories for photonic qubits are currently limited to a few milliseconds \CiteLongMemories. 
Here we report on a qubit memory based on a single atom coupled to a high-finesse optical resonator. 
By mapping and remapping the qubit between a basis used for light-matter interfacing and a basis which is less susceptible to decoherence, a coherence time exceeding \unit[100]{ms} has been measured with a time-independant storage-and-retrieval efficiency of 22\%.
This demonstrates the first photonic qubit memory with a coherence time that exceeds the lower bound needed for teleporting qubits in a global quantum internet.
}}

\indent 
Photons are convenient carriers for encoding both classical and quantum information.
To transport a pulse of light between the most distant locations on earth, it has to travel about \unit[20,000]{km}, which takes at least \unit[66]{ms}.
For future quantum networks allowing for distributed quantum computations, the exchange of quantum states between network nodes is indispensable.
In principle, it can be achieved by using single photons, but unavoidable losses in glass fibres in combination with the impossibility to amplify a quantum state renders the direct distribution of quantum states over long distances extremely inefficient.
A possible solution is the teleportation of qubits between the end nodes of a quantum repeater link \cite{kimble2008quantum,PhysRevLett.81.5932}.
While teleportation can be done deterministically, another challenge occurs:
as quantum teleportation requires classical communication between the end nodes, the receiver has to preserve its quantum state for at least the time it takes for the classical information to arrive \cite{bennet1992}.
This establishes the above-mentioned \unit[66]{ms} as a minimum requirement for global quantum-state distribution.
\newline
\indent
A qubit memory for photons has to combine a light-matter interface used to efficiently map the qubit between the photon and a material system with the ability to preserve quantum coherence for the duration of storage.
It has been shown that neutral atoms are suitable systems for both requirements \cite{treutlein2004coherence,specht}. 
Among the different possible atomic systems, single atoms offer a considerable advantage:
due to the reduced number of degrees of freedom, a single atom does not suffer from collective decoherence effects.
However, the internal states used for light-matter interfacing and maintaining coherence are usually different.
\figureOne

\indent
Here we present a solution by combining the advantages of two internal qubit configurations enabling us to reach a new regime of photonic qubit storage.
We utilise a previously demonstrated light-matter interface \cite{specht} to efficiently map the polarisation state of a single photon onto a pair of atomic eigenstates $\left\{\QubitState \right\}$ which we refer to as the \emph{\IB} \BFigure{puzzlepic}{a}.
As for many cold atoms experiments, the coherence of the atomic superposition is mainly degraded by magnetic field fluctuations.
Instead of reducing those \cite{ruster2016long,riedl2012bose} by e.g. employing a $\mu$-metal shielding, we resolve the principal problem by mapping the qubit between the \IB and a \emph{\SB} $\{\Fmf{1}{{-}1}, \Fmf{2}{{+}1}\}$ to preserve coherence during storage \BFigure{puzzlepic}{b}.
Since the differential weak-field magnetic susceptibility of this memory basis is $504.8$ times smaller compared to the \IB , a significant increase in coherence time is observed and further extended by the use of a spin-echo technique \BFigure{puzzlepic}{c}.

Our light-matter interface is based on a single \Rb atom trapped in a high-finesse optical resonator with $(g,\kappa,\gamma) = 2\pi \unit[(4.9, 2.8, 3.0)]{MHz}$. 
Here, the light-matter coupling rate on the relevant $\Fmf{1}{0} \to \FPmf{1}{1}$ transition is given by $g$, and the decay rates of the cavity field and the atomic dipole are given by $\kappa$ and $\gamma$, respectively.
We deterministically prepare a single atom in a far red-detuned dipole trap (\unit[1064]{nm}) at the centre of the cavity \cite{neuzner2016interference}.
In addition, a blue-detuned intracavity trap has been implemented to strongly confine the atom in the direction of the cavity axis to an antinode of the cavity mode resonant with the incoming photons, thus maximising the atom-cavity coupling $g$ \BFigure{schematics}{a}.
\figureTwo
The flying qubit is mapped onto and retrieved from the atom by means of a stimulated Raman adiabatic passage (STIRAP) \cite{specht} \sup{I}:
\begin{align*}
\begin{split}
 &\alpha \ket{\sigma^+} + \beta e^{\iu \phi} \ket{\sigma^-} \longleftrightarrow \alpha \ket{m_F{=}{-}1}+
 \beta e^{\iu \phi} \ket{m_F{=}{+}1} \\
 &\text{with }|\alpha|^2+|\beta|^2 =1, \text{ and } \alpha,\beta,\phi \in \mathbb{R}\,,
 \end{split}
\end{align*}
where $\sigma^{\pm}$ denote the circular polarisations \BFigure{schematics}{b}. 
The incoming qubit consists of a weak coherent laser pulse resonant with the cavity and containing one photon on average.
We define the efficiency of the memory as the probability to retrieve a photon when initially there was a single photon in the free space mode coupled to the cavity.
The efficiency of our system is measured to be 22\% \sup{III}.
Note that in contrast to our system, quantum memories based on atomic ensembles suffer from decaying efficiencies due to the decoherence of the collective spin wave \cite{bao2012efficient,zhao2009millisecond,yang2016efficient}.

To characterise the deviation from an ideal qubit memory, we determine the fidelity 
$\Fidelity {=} \bra{\psi_{\text{in}}} \hat{\rho}_{\text{out}} \ket{\psi_{\text{in}}}$ 
between the input state 
$\ket{\psi_{\text{in}}}$ 
and the output state 
$\hat{\rho}_{\text{out}}$.
To this end, quantum state tomography is performed as a function of storage time on six input polarisations in three mutually unbiased bases.

In an elementary store-and-retrieve experiment where only the \IB is used for storage \BFigure{cstplot}{ yellow crosses} as performed in \cite{specht}, the fidelity decays within hundreds of microseconds.
A magnetic guiding field (\unit[44]{mG}) used to reduce decoherence due to magnetic field fluctuations perpendicular to the guiding field \cite{specht} allows the extension of the coherence time by a factor of two but
induces a Larmor precession for linear polarisation inputs allowing only for discrete readout times. 
To become less sensitive to magnetic field fluctuations the qubit is 
remapped from the \IB to the \SB.
To this end, the population in $\Fmf{2}{{-}1}$ is transferred to $\Fmf{1}{{-}1}$ via a stimulated Raman transition (SRT) close to the \DOne line \BFigure{schematics}{c} \sup{II}.
\figureThree
The transfer, which lasts $\unit[40]{\mu s}$, is done immediately after the writing process and is reversed before the readout.
The mapping and remapping processes do not only require the transfer of the population from one state to another, but also the control of the qubit phase. 
First, the SRT will imprint the phase difference of the Raman pair onto the qubit. 
Second, once transferred to the \SB, the qubit rotates with $\Delta_\text{hf} \approx \unit[6.8]{GHz}$ due to the hyperfine splitting of the \Rb ground levels. 
Hence, by tuning the frequency difference of the Raman pair to the same frequency as the rotation of the qubit, the remapping SRT will compensate both: 
the rotation of the qubit during the storage and the initial phase of the first SRT. 
Therefore, it is possible to readout the qubit at any time.

Figure \ref{cstplot} shows the measured fidelity which beats the classical limit of $\Fidelity{=}\, 2\slash 3$ \cite{specht} for storage times longer than \unit[10]{ms}, which is almost two orders of magnitude longer compared to coherence times achieved without the mapping to the \SB.
As for the case of storing in the interface basis, circularly-polarised photons are mapped onto a single energy eigenstate and are not affected by fluctuations of the energy difference. 
From the fidelities at shortest storage times, one can notice that the fidelity is reduced by ($5{\pm}2$)\%.
This is due to off-resonant scattering during the optical SRT pulses limiting the maximum fidelity achievable with this optical transfer.

Theoretically, the magnetic-field sensitivity is reduced by a factor of $504.8$ when a qubit is stored in the \SB.
This predicts an increase in coherence time from \unit[200]{\textmu s} to \unit[100]{ms}, which is much longer than the observed \unit[10]{ms}.
We attribute the discrepancy to two mechanisms: 
first, for the qubit in the \SB, the motional degrees of freedom become important, because the trap frequencies for an atom in \ket{F{=}1} and \ket{F{=}2} are not identical.
This results in a motional-state-dependant energy difference and thus a dephasing of the qubit over time.
For our experimental parameters, we evaluated a trap frequency difference of $\delta \omega = \omega_{F=2}{-}\omega_{F{=}1} = 2\pi \unit[(13.7,\,264,\,0.2)]{Hz}$ for the $(x, y, z)$ axis. 
To avoid decoherence due to $\delta \omega_y$ we cool to the motional ground state along this axis during the initialization by means of Raman sideband cooling \cite{Reiserer2013,Boozer2006}.
However, the system is not cooled to the motional ground state along the $x$-axis. This attributes to the current limitation of \unit[10]{ms} coherence time.

A second limitation comes from the fact that transversely to the red-detuned trap axis the differential light-shift is a function of the position of the atom:
along $z$, only one potential well exists with its centre on the cavity axis.
Therefore, the atom is confined at the maximum of intensity, resulting in a known differential light shift.
In contrast, along $y$ (the axis of the blue-detuned standing-wave trap) different atoms may occupy different lattice sites.
Each of them is associated to a given differential light shift due to the change of the red-detuned trap intensity along the $y$-axis.
This spatially varying level splitting contributes to decoherence when averaged over many atoms.
By using an imaging system we post-select on the position of the atoms, but we are limited by spatial drifts of the traps and the imaging system.

\figureFour

Those two mechanisms result in a varying energy difference which is constant during one storage attempt. 
Therefore, these can be compensated by using a spin-echo technique, swapping the populations of the \SB states at half-storage-time \BFigure{puzzlepic}{c}.
To drive the $\Delta m_F=2$ transition, we use a Raman pair with a linear polarisation orthogonal to the cavity axis ($\sigma^+/\sigma^-$-polarised light) \BFigure{schematics}{c}.
Again, the relative phase of the Raman pair has to be precisely controlled.

Figure \ref{seplot} shows that the combination of storage in the \SB with a single spin-echo pulse \BFigure{puzzlepic}{c} increases the coherence times to more than \unit[100]{ms}.
As for the transfer to the \SB, the spin-echo pulse degrades the fidelity by ($5\pm2$)\% due to off-resonant scattering on the \DOne line. 
With this extended storage time, a new source of decoherence must be taken into account:
scattering from the dipole traps leads to dephasing and also to mixing of populations.
Indeed, we observe \BFigure{seplot}{} that the fidelity of both, linear and circular input polarisations, decays over time.
The red as well as the blue-detuned trap contribute to this effect. 
A theoretical estimation gives a scattering rate of \unit[2]{Hz} \sup{IV}. This number agrees with the observed decay of the fidelity \BFigure{seplot}{}.
The efficiency of the presented qubit memory is governed by the parameters of our optical resonator \cite{dilley2012single}. 
Increasing the cooperativity by choosing a smaller mode volume or using higher-quality mirrors with lower intracavity losses will further improve the efficiency.
All the discussed limitations of the storage time and fidelity are not fundamental and can be overcome technologically or by implementing additional experimental capabilities: 
for instance, 
additional Raman sideband cooling  
in order to reach the motional ground state along the $x$-axis. 
The loss in starting fidelity caused by the optical Raman transfers can be avoided by using microwave transitions instead.
However, the current design of our apparatus does not allow for a sufficiently high Rabi frequency for a reasonable microwave power.
The light shift, as well as the off-resonant scattering from the dipole traps, can be reduced by lowering the power and/or increasing the detuning. 
\newline
\indent
In summary, we have presented a memory for a photonic qubit using a combination of two different atomic configurations.
This leads to an increase in coherence time by three orders of magnitude compared to prior work \cite{specht} and pushes the current state-of-the-art by two orders of magnitude.
Hence, this neutral atom based cavity system opens up new possibilities for the implementation of novel quantum communication protocols including quantum repeaters \cite{razavi2009quantum,uphoff2016integrated} where memories are essential ingredients.

\vbox{
\noindent\textbf{Acknowledgments} \\
We thank B. Wang for the development of hardware and M. Uphoff and S. Dürr for discussions.
This work was supported by the Bundesministerium für Bildung und Forschung (BMBF, Verbund Q.com - Q) and by the Deutsche Forschungsgemeinschaft via the excellence cluster Nanosystems Initiative Munich (NIM).

\bigskip

\noindent\textbf{Author contributions} \\
M.K., O.M., A.N., S.R. and G.R. conceived the experiment.
M.K., O.M. and S.L. performed the experiment.
M.K., O.M., S.L., S.R. and G.R. evaluated the data.
All authors contributed to the writing of the manuscript.

\bigskip

\noindent\textbf{Competing financial interests} \\
The authors declare no competing financial interests.
}

\end{document}